\pdfoutput=1
\documentclass[sigconf]{acmart}
\usepackage{xcolor}
\usepackage{colortbl}
\usepackage{tcolorbox}
\usepackage{listings}
\usepackage{makecell}
\usepackage{multirow}
\usepackage{amsmath,amsfonts}

\usepackage{amssymb}
\usepackage{hyperref} 
\usepackage{algorithmic}
\usepackage{graphicx}
\usepackage{fbox}
\usepackage{textcomp}
\usepackage{minibox}
\usepackage{balance}
\usepackage{enumitem}
\usepackage{xspace}
\usepackage[T1]{fontenc}
\usepackage{booktabs}
\usepackage{threeparttable}
\usepackage[]{collab}
\usepackage{url}
\usepackage{natbib}

\makeatletter

\newcommand{\Rmnum}[1]{\expandafter\@slowromancap\romannumeral #1@}
\makeatother

\definecolor{codegreen}{rgb}{0,0.6,0}
\definecolor{codegray}{rgb}{0.5,0.5,0.5}
\definecolor{codepurple}{rgb}{0.58,0,0.82}
\definecolor{backcolour}{rgb}{0.95,0.95,0.92}

\lstdefinestyle{mystyle}{
    backgroundcolor=\color{backcolour},   
    commentstyle=\color{codegreen},
    keywordstyle=\color{magenta},
    numberstyle=\tiny\color{codegray},
    stringstyle=\color{codepurple},
    basicstyle=\ttfamily\footnotesize,
    breakatwhitespace=false,         
    breaklines=true,                 
    captionpos=b,                    
    keepspaces=true,                 
    numbers=left,                    
    numbersep=5pt,                  
    showspaces=false,                
    showstringspaces=false,
    showtabs=false,                  
    tabsize=2
}

\AtBeginDocument{%
  \providecommand\BibTeX{{%
    \normalfont B\kern-0.5em{\scshape i\kern-0.25em b}\kern-0.8em\TeX}}}

\copyrightyear{2024} 
\acmYear{2024} 
\setcopyright{acmlicensed}\acmConference[ICSE-SEIP '24]{46th International Conference on Software Engineering: Software Engineering in Practice}{April 14--20, 2024}{Lisbon, Portugal}
\acmBooktitle{46th International Conference on Software Engineering: Software Engineering in Practice (ICSE-SEIP '24), April 14--20, 2024, Lisbon, Portugal}
\acmDOI{10.1145/3639477.3639754}
\acmISBN{979-8-4007-0501-4/24/04}

\newcommand\cloud{CloudA\xspace}
\newcommand\nm{FaultProfIT\xspace}

\newcommand{\etc}{{\em etc}\xspace}
\newcommand{\ie}{{\em i.e.},\xspace}
\newcommand{\eg}{{\em e.g.},\xspace}

\definecolor{ballblue}{rgb}{0.13, 0.67, 0.8}
\definecolor{jcpink}{RGB}{255, 0, 96}
\collabAuthor{jy}{ballblue}{Jinyang Liu}
\collabAuthor{jc}{jcpink}{JC}
\collabAuthor{jz}{magenta}{Jiazhen Gu}
\collabAuthor{zh}{green}{Zhihan Jiang}
\collabAuthor{jj}{purple}{Junjie}
\collabAuthor{zb}{teal}{Zhuangbin}
\collabAuthor{yc}{blue}{Yichen}

\begin{document}

\title{FaultProfIT: Hierarchical Fault Profiling of Incident Tickets in Large-scale Cloud Systems}

\author{Junjie Huang}
\affiliation{%
  \institution{The Chinese University of Hong Kong}
  \city{Hong Kong}
  \country{China}}

\author{Jinyang Liu}
\affiliation{%
  \institution{The Chinese University of Hong Kong}
  \city{Hong Kong}
  \country{China}}
\author{Zhuangbin Chen}
\affiliation{%
  \institution{Sun Yat-sen University} \country{China}}

\author{Zhihan Jiang}
\affiliation{%
  \institution{The Chinese University of Hong Kong}
  \city{Hong Kong}
  \country{China}}

\author{Yichen Li}
\affiliation{%
  \institution{The Chinese University of Hong Kong}
  \city{Hong Kong}
  \country{China}}

\author{Jiazhen Gu}
\authornote{Corresponding author.}
\affiliation{%
  \institution{The Chinese University of Hong Kong}
  \city{Hong Kong}
  \country{China}}

\author{Cong Feng}
\author{Zengyin Yang}
\affiliation{%
  \institution{Computing and Networking Innovation Lab, Huawei Cloud Computing Technology Co., Ltd}
  \country{China}}
  
\author{Yongqiang Yang}
\affiliation{%
  \institution{Computing and Networking Innovation Lab, Huawei Cloud Computing Technology Co., Ltd}
  \country{China}}

\author{Michael R. Lyu}
\affiliation{%
  \institution{The Chinese University of Hong Kong}
  \city{Hong Kong}
  \country{China}}

\renewcommand{\shortauthors}{Junjie Huang, et al.}

\begin{abstract}
Postmortem analysis is essential in the management of  incidents within cloud systems, which provides valuable insights to improve system's reliability and robustness.
At \cloud\footnote{Due to the company policy, we anonymize the name as \cloud.}, \textit{fault pattern profiling} is performed during the postmortem phase, which involves the classification of incidents' faults into unique categories, referred to as \textit{fault pattern}.
By aggregating and analyzing these fault patterns, engineers can discern common faults, vulnerable components and emerging fault trends.
However, this process is currently conducted by manual labeling, which has inherent drawbacks. 
On the one hand, the sheer volume of incidents means only the most severe ones are analyzed, causing a skewed overview of fault patterns.
On the other hand, the complexity of the task demands extensive domain knowledge, which leads to errors and inconsistencies.

To address these limitations, we propose an automated approach, named \nm, for \textbf{Fault} pattern \textbf{Prof}iling of \textbf{I}ncident \textbf{T}ickets.
It leverages hierarchy-guided contrastive learning to train a hierarchy-aware incident encoder and predicts fault patterns with enhanced incident representations. 
We evaluate \nm using the production incidents from \cloud. The results demonstrate that \nm outperforms state-of-the-art methods. Our ablation study and analysis also verify the effectiveness of hierarchy-guided contrastive learning. 
Additionally, we have deployed \nm at \cloud for six months. To date, \nm has analyzed 10,000+ incidents from 30+ cloud services, successfully revealing several fault trends that have informed system improvements.

\end{abstract}

\ccsdesc[500]{Software and its engineering~Software evolution}
\ccsdesc[500]{Software and its engineering~Maintaining software}

\keywords{Incident management, incident tickets, fault patterns}

\maketitle
\section{Introduction}\label{sec:intro}

\begin{figure}[t]
    \centering
    \includegraphics[width=0.94\columnwidth]{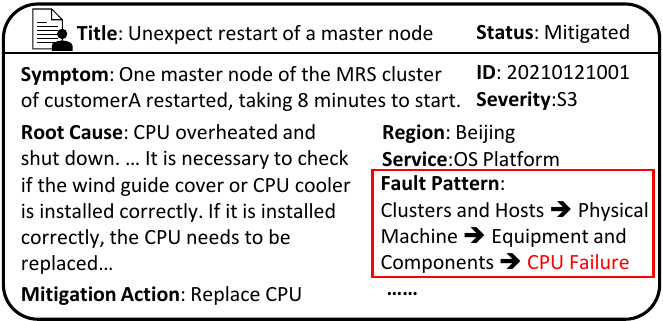}
    \vspace{-0.3em}
    \caption{An example of an incident ticket.}
    \label{fig:incident-ticket}
    \vspace{-1.7em}
\end{figure}

Production incidents, which represent unplanned service interruptions or performance degradation, are inevitable in large-scale cloud services~\cite{chen2020towards,wang2021howlong}. 
They could decrease customer satisfaction and cause huge economic losses~\cite{chen2020towards,facebook2021outage,li2021fogofwar}.
To effectively manage these incidents, cloud vendors (\eg Amazon Web Service~\cite{aws}, Microsoft Azure~\cite{azure}, and Google Cloud Platform~\cite{google}) have developed incident management systems~\cite{ghosh2022fightsocc} for prompt incident detection, diagnosis, and resolution.
In such systems, the details of an incident are typically documented in an \textit{incident ticket} (see an example in Figure~\ref{fig:incident-ticket}), including its title, symptom, and resolution status, \etc.
The ticket is then tracked and updated throughout the incident lifecycle until the issue is resolved~\cite{chen2020towards}.
In general, the entire lifecycle of an incident can be divided into two main phases, \ie \textit{real-time response} and \textit{postmortem analysis}~\cite{ghosh2022fightsocc}.
The former aims to quickly mitigate the incident's impact upon its occurrence.
After incident resolution, the latter retrospectively examines the tickets to gain valuable insights that can enhance future incident management~\cite{lou2013software, wang2021howlong}.

Postmortem analysis plays an essential role in the continuous improvement of cloud systems' reliability and robustness~\cite{gunawi2016does, ghosh2022fightsocc}. 
Specifically, it is conducted to understand the root cause of the incident, assess the impact, and evaluate the mitigation process, which can potentially be used to prevent similar incidents from happening again.
Existing studies have demonstrated that production incidents could be recurring~\cite{shetty2022softner,li2022actionable} or share certain similarities~\cite{chen2020identifying}.
Thus, the knowledge and insights derived from the postmortem analysis often exhibit recurring patterns.
By aggregating and categorizing such recurring patterns, we can identify common faults, solutions, vulnerable components, and trends in the large volume of incidents, which can serve as a reference guide for understanding, diagnosing, and resolving future incidents more efficiently.

\begin{figure}[t]
    \centering
    \includegraphics[width=0.94\columnwidth]{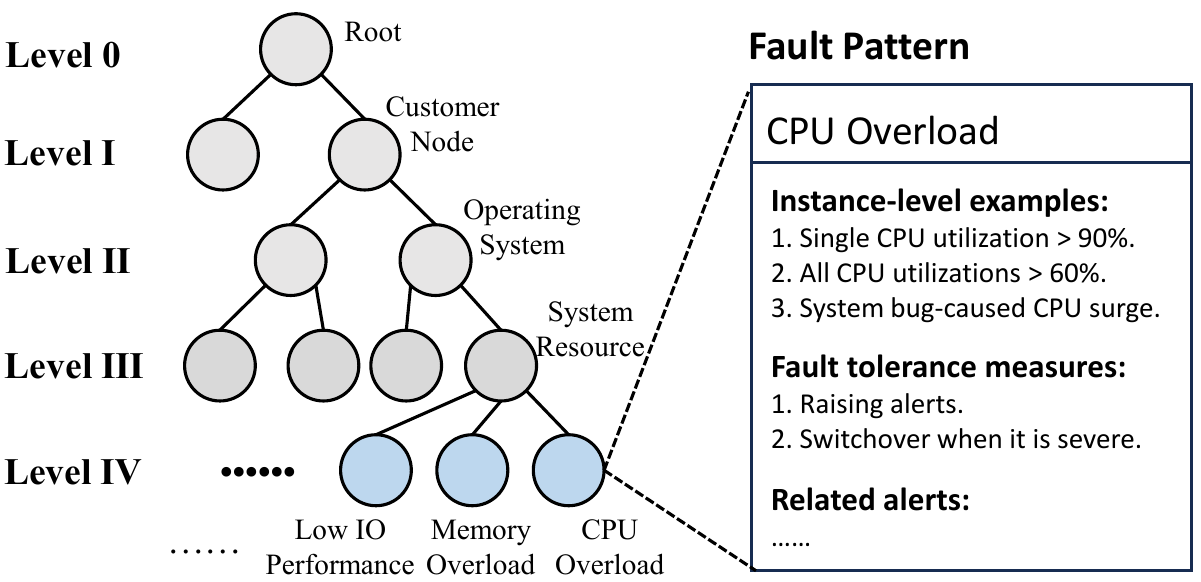}
    \vspace{-0.5em}
    \caption{Fault pattern example in the hierarchical taxonomy.}
    \label{fig:fault-pattern-tree}
    \vspace{-1em}
\end{figure}

Based on this fact, the reliability engineers of \cloud (a top-tier cloud vendor offering global online services) perform the task of \textit{fault pattern profiling} during postmortem analysis.
This task involves classifying the faults that occurred during incidents into distinct categories, such as CPU overload, power outage, SSD failure, \etc.
We refer to each category as a \textit{fault pattern}, which is a concise representation of the fault, including a fault name, a set of typical phenomena seen in historical examples, a list of possible mitigation measures, \etc.
A fault pattern of CPU failure is illustrated in Figure~\ref{fig:incident-ticket}, which describes the breakdown of a physical machine in a cluster due to overheating.
Replacing the CPU mitigates the issue.
Clearly, the fault pattern offers readily information about the symptom and root cause of the incident together with actionable suggestions, which significantly accelerates the incident management pipeline.
Due to the large scale and complexity of cloud systems, there exist a tremendous number of fault patterns.
To better manage and exploit this knowledge,  they are organized as a tree-like taxonomy based on their position across the entire cloud system stack (see Section~\ref{sec:fault-pattern-profiling}).
Figure~\ref{fig:fault-pattern-tree} presents an example of such hierarchical taxonomy with five levels, which shows a customer node suffering a CPU overload issue.
At \cloud, engineers value the role of fault patterns and have accumulated 334 of them.

In current practice, fault pattern profiling is carried out manually.
This procedure involves carefully examining the tickets, identifying useful information, and aligning with the fault pattern taxonomy.
While the manual approach is effective, it is time-consuming and prone to human error.
First, the overwhelming volume of incidents implies that only a small fraction of incidents can be selected for in-depth postmortem analysis.
This sampling may result in fault patterns that reflect only a partial distribution, thereby leading to a skewed overview and potentially sub-optimal improvement decisions. 
Second, the inherent complexity of the labeling task demands a deep understanding of the entire fault pattern hierarchy and the nature of incidents.
Such requirements go beyond the ability of a single engineer.
Thus, manual labeling will inevitably introduce errors and inconsistencies, leading to a distorted fault pattern distribution.
Furthermore, the taxonomy of fault patterns is not static; it continuously evolves with the introduction of new patterns and adjustments in existing hierarchical relations.
As a result, it is imperative to develop an automated approach for fault pattern profiling that can accommodate these complexities.

However, training a model capable of learning from existing fault patterns to automatically profile the unseen incident tickets is a non-trivial task, which presents the following two major challenges. 
First, fault patterns possess rich and complicated information, making their features hard to be exploited.
As shown in Figure~\ref{fig:fault-pattern-tree}, each fault pattern not only has a structural position in the hierarchy but also includes explicit textual descriptions.
The challenge lies in effectively harnessing such hybrid features, both hierarchical and textual, to accurately predict the fault patterns for unseen tickets.
Second, the limited size of training samples poses another problem.
The development of a robust fault pattern profiling model requires a substantial volume of labeled incident tickets.
However, manually profiling fault patterns is both expensive and error-prone.
Thus, only limited labeled examples are produced during the daily service maintenance at \cloud.

To address these challenges, we propose \nm, an automatic approach for \textbf{Fault} patterns \textbf{Prof}iling of \textbf{I}ncident \textbf{T}ickets based on hierarchical textual classification (HTC)~\cite{zangari2023htc_survey}.
Specifically, we employ hierarchy-guided contrastive learning~\cite{wang2022hgclr} to train a hierarchical text classifier, aiming to precisely encode the sophisticated features of fault patterns while addressing the problem of data insufficiency.
Contrastive learning has long been recognized as an effective way to learn meaningful textual representations~\cite{kim2021self, rethmeier2023clr_survey} with limited training samples by augmenting positive and negative samples and distinguishing among them~\cite{rethmeier2023clr_survey}. 
By concentrating on similar input samples and pushing apart dissimilar ones,
contrastive learning can enhance text representations and improve classification accuracy. 
In addition, to fully utilize the knowledge of fault patterns, we expand conventional contrastive learning to produce hierarchy-aware text representation.
We apply an optimized Graphormer~\cite{ying2021graphormer}, a powerful graph representation model based on Transformer layers~\cite{vaswani2017attention}, to encode the hierarchical structures and node descriptions together.
These representations capture both the semantics and hierarchy of fault patterns, and thus can support more accurate profiling. 

We have evaluated and deployed \nm at \cloud. Our evaluation demonstrates that \nm achieves a high degree of accuracy (78.3\% F1-score) in automatic profiling of fault patterns, outperforming a wide range of text classification models. We also conduct a comprehensive ablation study and analysis to demonstrate the effectiveness of our hierarchy-guided contrastive learning approach in learning hierarchy-aware incident representation. Furthermore, we profile fault patterns across incidents of various categories and find that some incidents, such as those associated with lighter severity and those from infrastructure and computing services, exhibit higher accuracy. Lastly, we have deployed \nm to a cloud reliability analysis system at \cloud, an incident management and analytics platform used by over 30 service teams and thousands of engineers. \nm has been running at \cloud for six months, and successfully identified a number of emerging fault trends, which in turn guide engineers to fix the vulnerabilities, thus improving system reliability.

To sum up, this paper makes the following contributions:
\begin{itemize}[leftmargin=*, topsep=0pt]
    \item To the best of our knowledge, we are the first to automatically profile fault patterns of incident tickets for postmortem analysis. 
    \item We propose \nm, which leverages hierarchy-guided contrastive learning to learn hierarchy-aware incident representation to classify pattern pattern labels.
    \item We conduct an extensive evaluation on the production incidents at \cloud. The results show that \nm outperforms state-of-the-art methods.
    \item We have deployed \nm at \cloud for six months, where it has analyzed over 10,000 incidents from 30+ cloud services and revealed fault trends for system improvements.
\end{itemize}

\begin{table*}[h]
  \centering
    \caption{Categories of fault patterns at level \uppercase\expandafter{\romannumeral1}.}
\vspace{-1mm}
    \resizebox{0.96\textwidth}{!}{
    \begin{tabular}{lp{10cm}p{6cm}}
    \toprule
  \multicolumn{1}{l}{\textbf{Level \uppercase\expandafter{\romannumeral1}}} & \multicolumn{1}{l}{\textbf{Description}} & \multicolumn{1}{l}{\textbf{Fault Pattern Example}}   \\
    \midrule
     \multirow{3}{*}{Infrastructure and Sites} &  {Incidents that occur within the physical and network infrastructure of a site.} This includes problems related to external and dedicated networks, data center environment and facilities, and data center network equipment.  & Infrastructure and Sites $\rightarrow$ Data Center Environment $\rightarrow$ Data Center Facilities $\rightarrow$ Power Supply Insufficient \\
     \cmidrule(lr){1-3}
     \multirow{2}{*}{Clusters and Hosts} & {Incidents that occur within the system clusters and specific hosts}, including the physical machines, virtual machines, containers, and storage.  & Clusters and Hosts $\rightarrow$ Physical Machine $\rightarrow$ Device and Components $\rightarrow$ SSD Failure \\
     \cmidrule(lr){1-3}
     \multirow{2}{*}{Customer Node} & {Incidents that specifically occur within the business nodes of customers}, impacting the operating system and its processes or threads. & Customer Node $\rightarrow$ Operating System $\rightarrow$ System Resources $\rightarrow$ CPU Overload \\
     \cmidrule(lr){1-3}
     \multirow{3}{*}{Load and Capacity} & {Incidents disrupt the balance and efficiency of system load and capacity management}, leading to performance degradation, traffic surges, security threats, and resource allocation issues.  & Load and Capacity $\rightarrow$ Overload Control $\rightarrow$ Frontend Load $\rightarrow$ Security Attacks \\
     \cmidrule(lr){1-3}
     \multirow{3}{*}{Business and Data} & Incidents that occur during the operation and management of business processes and data, including problems with tenant resources, business configurations, licensing, security credentials, and system configurations.  &  Business and Data $\rightarrow$ Tenant Resources $\rightarrow$ Tenant Account and Data $\rightarrow$ Tenant Data Deletion \\
     \cmidrule(lr){1-3}
     \multirow{3}{*}{Dependencies} & Incidents that occur within the internal and external dependencies of the system, including problems with web servers, databases, microservices, and cloud service dependencies. & Dependencies $\rightarrow$ Internal Dependency $\rightarrow$ Database $\rightarrow$ Unavailable Database Service \\
     \cmidrule(lr){1-3}
     \multirow{2}{*}{Disaster Recovery} & System's inability to recover and resume normal operations after a significant disruption over regions or availability zones. & Disaster Recovery $\rightarrow$ AZ Disaster Recovery $\rightarrow$ AZ Site $\rightarrow$ Active AZ Site Failure \\
    \bottomrule
    \end{tabular}%

}
  \label{tab:background-fp-first-level}%
\end{table*}

\section{Background and Motivation}
\subsection{Incident and Incident Management}
In cloud systems, an incident is defined as an unplanned interruption or performance degradation of a service or product that impacts service availability and customer satisfaction~\cite{ghosh2022fightsocc, chen2020towards}. 
For example, a slow connection, an unavailable service, and a customer-reported error could constitute an incident.

\subsubsection{Incident Lifecycle}
\begin{figure}[t]
    \centering
    \includegraphics[width=0.94\columnwidth]{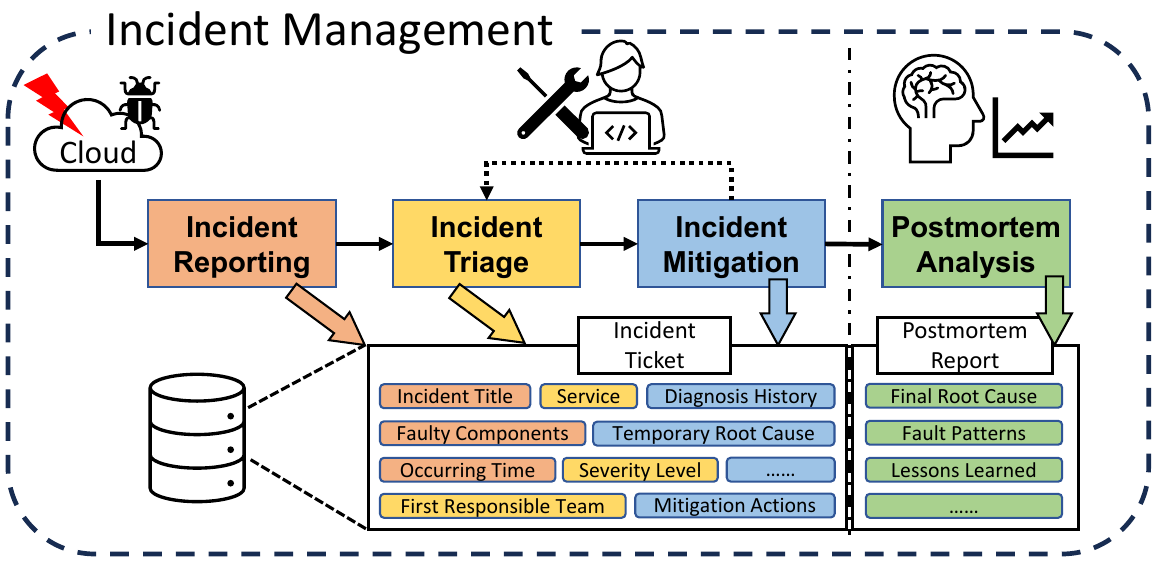}
    \vspace{-0.5em}
    \caption{The incident management process.}
    \label{fig:inci-life-cycle}
    \vspace{-1em}
\end{figure}

In order to accelerate incident mitigation and prevent the incident from happening again, cloud vendors such as \cloud build \textit{incident management} systems to assist engineers during the whole incident life cycle~\cite{chen2020towards, shetty2022softner, zhao2023manage}. Figure~\ref{fig:inci-life-cycle} shows a typical example of the incident lifecycle, which can be broadly divided into two phrases, \ie \textit{real-time response} and \textit{postmortem analysis}.

\noindent \textbf{Real-time Response.} When an incident occurs, On-Call Engineers (OCEs) must take immediate action to resolve the incident to minimize its impact.
This process begins with \textit{incident reporting}, where an incident is initially raised~\cite{li2021fogofwar,gu2020efficient}. 
In cloud systems, the incidents can be reported by customers when they encounter problems during service usage or detected by tailored system monitors when performance metrics fall below pre-defined acceptance levels. A severity level is also assessed to measure the impacts of each incident and determine whether additional investigation is required~\cite{ahmed2023knowledge}.
Then, in the \textit{incident triage} stage, the incident will be routed to an appropriate service team for resolution~\cite{chen2019triage}.
Based on service ownership and heuristic algorithms, the responsible team is automatically determined. However, due to the high complexity and dependencies, the incident could be incorrectly triaged. 
In this case, it will be re-routed to a more appropriate team for investigation, and this process can repeat several times~\cite{gao2020scouts}.
The final stage is \textit{incident mitigation}, in which the service team investigates the incident and takes mitigation actions to bring the problematic service back to normal. In practice, some temporary workarounds (\eg server rebooting and configuration change) will be applied first to quickly mitigate the impact. But occasionally, the team can encounter intractable problems. In such cases, they can escalate and engage additional teams for investigation, which often necessitates more complex mitigation measures to cloud systems, such as bug fixing and version rollback. 
Upon the resolution of the incident, it will be closed in the incident management system.
During the real-time response period, OCEs create incident reports as a record of diagnosis. The report is written by following some rigorous rules to ensure clarity, thereby facilitating subsequent diagnostic procedures and postmortem analyses. At \cloud, an incident report contains plentiful information, including an incident timeline, temporary root causes, escalating records, mitigation actions, \etc.

\noindent \textbf{Postmortem Analysis.}
At \cloud, when an incident is resolved, a \textit{postmortem analysis} will often be conducted to evaluate the circumstances retrospectively. 
The insights gleaned from this analysis are invaluable, serving as important guidelines for the continuous improvement and enhancement of the cloud system's reliability. 
During analysis, engineers need to write a postmortem report to reflect the whole incident picture and summarize useful knowledge for future retrospection. At \cloud, a postmortem report derives from the original incident ticket and the reliability team will involve more contents to it such as (1) the final root causes of incidents, (2) fault categories to tag the incident, and (3) suggestions to system improvement. Such contents are maintained in natural language and categorical data, which are easily accessible and shareable. 
Due to the large volume of incidents and limited human resources, only a small sample of incidents will be selected and analyzed during postmortem. The selection criteria of incidents are ad-hoc, mainly based on engineer feedback, incident severity, and observed trends. 
Our work applies directly to enhancing postmortem analysis and contributing to continuous improvement in cloud reliability, which deals with automatic fault profiling for incident tickets.

\subsubsection{Incident Management System}
At \cloud, thousands of incidents are reported to OCEs every day from various sources such as external customers, internal engineers, and automated monitoring systems. To manage the incidents at scale, \cloud developed a web application for company-wide incident reporting, investigation, and analysis. During an incident's lifecycle, an incident ticket is well-documented to record the incident-related information by various participants, such as OCEs to verify and report the incidents, SREs to mitigate the incidents, and cloud reliability teams to conduct postmortem analysis and derive valuable insights. 
Figure~\ref{fig:incident-ticket} shows an example of incident tickets. 
These tickets contain fruitful information such as incident timelines, severity, summaries, root causes, and mitigation suggestions written by SREs.

\subsection{Fault Pattern Profiling}\label{sec:fault-pattern-profiling}
\textit{Fault pattern profiling} is a crucial task employed by the reliability team at \cloud to derive insights from postmortem analysis. This task involves classifying the faults that occurred during incidents into distinct categories, which is referred to as \textit{fault pattern}. 

\noindent \textbf{Fault Pattern.} At \cloud, a fault pattern characterizes abnormal behaviors exhibited in specific objects. Each fault pattern comprises a fault name, a set of potential phenomena, measures for fault tolerance, \etc. An example of a fault pattern, as shown in the right segment of Figure~\ref{fig:fault-pattern-tree}, is \textit{CPU overload}. This fault pattern is described by phenomena such as a single CPU utilization exceeding $90\%$, and fault tolerance strategies such as switchover. The concise representation of faults as fault patterns enables engineers to readily understand the nature of the faults, thereby facilitating efficient fault diagnosis and mitigation.

\noindent \textbf{Fault Pattern Taxonomy.} The reliability team has developed a comprehensive \textit{fault pattern taxonomy} to manage a multitude of fault patterns across diverse objects. This taxonomy is structured in a tree-like hierarchy with five levels, comprising 7 hyper classes at level \uppercase\expandafter{\romannumeral1} and 334 fault patterns at leaf nodes. For example, as shown in the left segment of Figure~\ref{fig:fault-pattern-tree}, a \textit{Customer Node} consists of the Operating System level and the Process level, which can be further subdivided into system resource, environment, and so on.

The principle of constructing the taxonomy is to divide and group fault patterns based on the specific components where the faults occur. In practice, it was initiated by analyzing historical incident records to identify common patterns. Similar fault phenomena in the same object were summarized into a single fault pattern. Subsequently, similar fault patterns were grouped into a hyper-class based on the specific objects in which they all occur. The hyper-class can broadly contain region-level or az level components, but it can also be narrowed down to a VM or a system environment. For example, CPU overload, memory overload, and low IO performance all reflect different aspects of system resources.
With the dedicated partitioning, the reliability produced the first version of the taxonomy, which has been maintained for over eight years and has undergone multiple rounds of refinement. It is now considered comprehensive and ready-to-use, and is continuously updated to accommodate new incidents and fault patterns. Table~\ref{tab:background-fp-first-level} shows the seven top-level categories in the taxonomy, each consisting of a brief description and multiple finer-grained subcategories covering faults in various system components.
The fault pattern taxonomy is a valuable knowledge base of great utility, which has also been successfully applied in other reliability scenarios at \cloud, such as guiding fault injection and informing disaster recovery design.

\noindent \textbf{Automatic Fault Pattern Profiling.}
A crucial aspect of postmortem analysis at \cloud involves the examination of incident fault patterns. The objective of this process is to categorize incidents for follow-up analysis. Figure~\ref{fig:incident-ticket} shows an incident and its associated fault pattern. By evaluating the distribution of fault patterns over a specific timeframe, the reliability team can discern system trends and recurring faults. This data-driven understanding can guide strategic business decisions and assist in setting overarching group targets. Additionally, engineers can leverage fault pattern categories to retrieve relevant incidents of a similar nature, thereby serving as a valuable reference during fault diagnosis and a knowledge repository for experience sharing. 

The current practice of fault pattern profiling is conducted through manual labeling during the postmortem phase. Reliability engineers first examine the diagnostic details within the tickets and then identify anomalous behaviors, which are typically indicated in natural language in the tickets. Subsequently, they align the information in fault patterns with the incident to determine fault patterns. Practically, a single incident can exhibit multiple fault patterns, indicating the simultaneous occurrence of multiple faults. 

\begin{figure}[t]
    \centering
    \includegraphics[width=0.9\columnwidth]{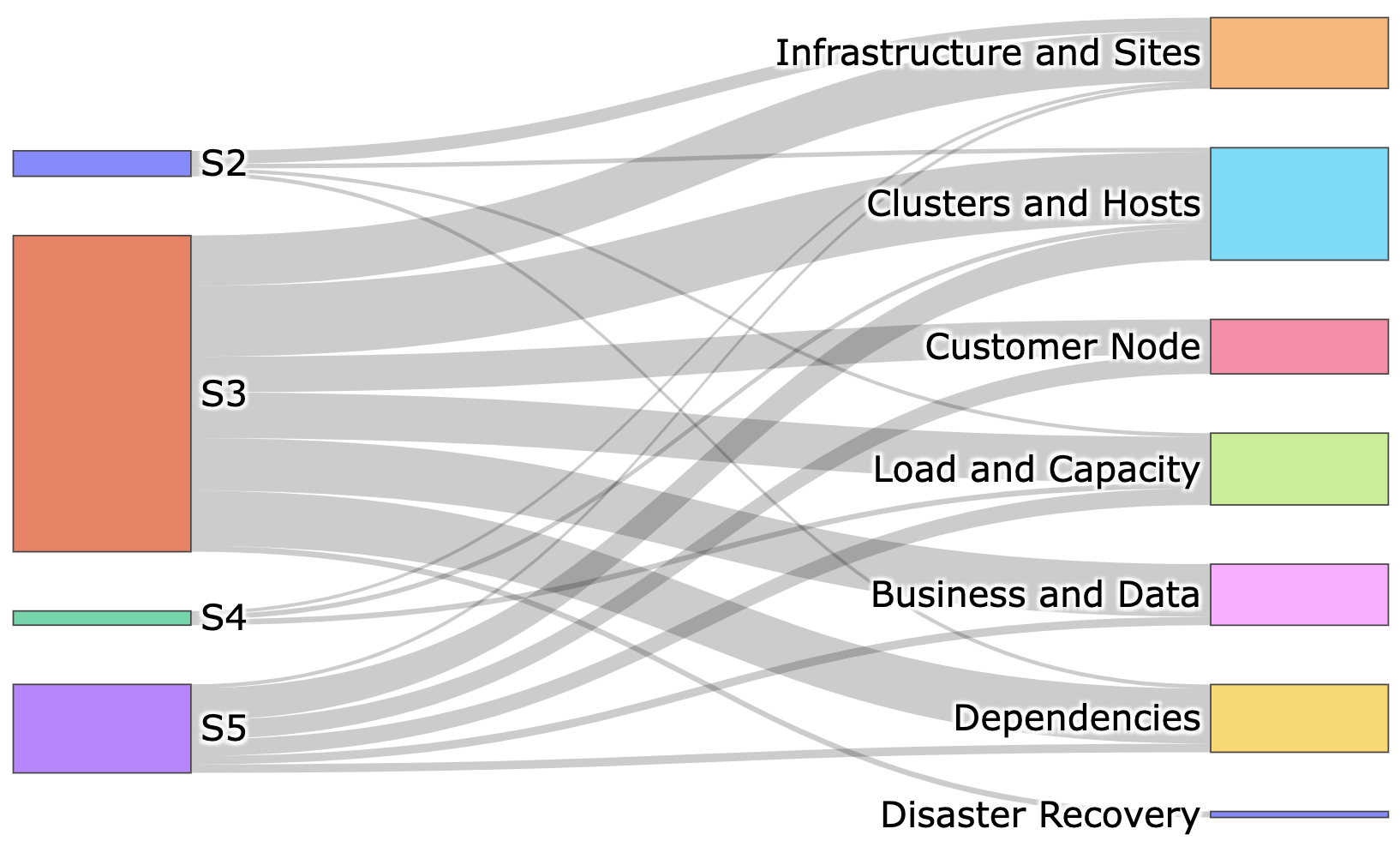}
    \vspace{-1em}
    \caption{Fault pattern distribution of incidents with different severities that have undergone postmortem.}
    \label{fig:sankey-fp-severity}
    \vspace{-1.4em}
\end{figure}

Despite the effectiveness of manual profiling, it is labor-intensive and susceptible to errors, potentially yielding biased insights regarding system maintenance. Firstly, the overwhelming volume of incidents means that only a small fraction of these incidents undergo postmortem analysis due to the limited human resources and high cost of analysis~\cite{dogga2023autoarts}. Incidents of high severity are prioritized, resulting in less critical ones being neglected, thereby making partial coverage. For example, Figure~\ref{fig:sankey-fp-severity} shows the fault pattern distribution with respect to incidents of different severity that has been analyzed during postmortem. The incidents with severity S3 significantly outnumber less severe ones. However, in reality, incidents in S4 and S5 should be more prevalent. Secondly, the assignment of fault pattern types requires extensive domain knowledge of incidents and fault patterns. However, the varying expertise levels of engineers can introduce errors and inconsistencies in manual profiling. For example, 29\% root cause tags assigned by OCEs at Microsoft are incorrect~\cite{dogga2023autoarts}. 

\begin{figure*}[t]
    \centering
    \includegraphics[width=0.92\textwidth]{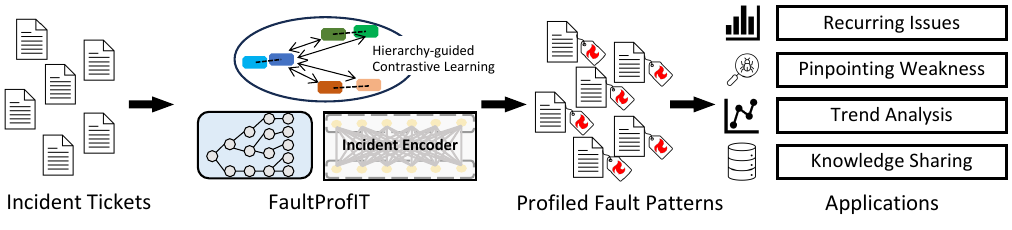}
    \vspace{-1em}
    \caption{The overview of automatic fault pattern profiling.}
    \label{fig:fault-pattern-profiling}
    \vspace{-0.3em}
\end{figure*}

To address these issues, our work introduces techniques to automate the profiling of fault patterns in incident tickets. Figure~\ref{fig:fault-pattern-profiling} shows an overview of the task. Our approach not only improves the efficiency of postmortem analysis but also provides more accurate fault patterns for downstream applications.

\section{Methodology}~\label{sec:method}
As discussed before, manually profiling fault patterns for incident tickets is labor-intensive and error-prone, leading to a biased understanding of faults in cloud systems. To address the issue, we propose \nm, an automated tool for fault pattern profiling at \cloud. \nm utilizes language models to read the diagnostic descriptions in incident tickets and predicts the fault pattern labels from the taxonomic hierarchy, which can improve the efficiency of postmortem analysis and provide actionable insights for business decision making. 
In this section, we introduce our method in detail.

\vspace{-1mm}
\subsection{Overview}
\begin{figure}[t]
    \centering
    \includegraphics[width=0.96\columnwidth]{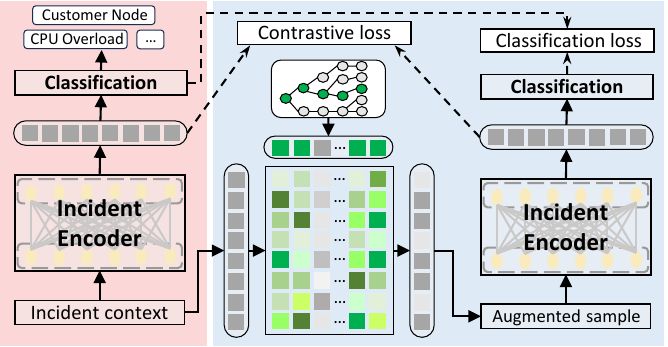}
    \caption{The overview of \nm. Parts in blue and red denote the training and prediction workflow, respectively.}
    \label{fig:overview-method}
    \vspace{-1em}
\end{figure}

\nm utilizes hierarchical text classification techniques~\cite{zangari2023htc_survey, wang2022hgclr} to predict fault patterns for incident tickets. The main concept of \nm is to employ pretrained language models (PLM)~\cite{devlin2019bert} to comprehend the semantics of incident tickets and incorporate a taxonomic hierarchy into the PLM to produce hierarchy-aware representations for classification. Figure~\ref{fig:overview-method} shows an overview of \nm. Given an incident ticket, we first extract relevant data from the ticket to establish the incident context~(\S~\ref{sec:method-preprocessing}). Subsequently, we apply an incident encoder based on the MacBERT PLM~\cite{cui2020macbert} to encode incident context into vector features for classification~(\S~\ref{sec:method-text-encoder}). The encoder is trained to incorporate hierarchical fault pattern information by the hierarchy-guided contrastive learning~\cite{wang2022hgclr}. In contrastive learning, building challenging positive samples is crucial~\cite{rethmeier2023clr_survey}. Therefore, guided by the taxonomic hierarchy~(\S~\ref{sec:method-hierarchy-encoder}), we construct high-quality positive samples that are both label-involved and hierarchy-aware for the incident context~(\S~\ref{sec:method-positive-data}). By pulling closer to the original incident contexts with augmented samples, the incident encoder can learn to generate hierarchy-aware textual representations~(\S~\ref{sec:method-clr}). Finally, after training, \nm can discard the redundant hierarchy and utilize the hierarchy-enhanced incident encoder to classify fault patterns.

\subsection{Incident Data Fetching and Preprocessing}\label{sec:method-preprocessing}
During the incident lifecycle, different groups of engineers collaboratively contribute to different fields of the incident tickets at different stages. In order to help the postmortem process and prevent any data leakage, we assume only the fields of tickets before postmortem analysis can be available for fault pattern profiling (Figure~\ref{fig:inci-life-cycle} shows the typical fields before postmortem). In our work, we select four types of information from tickets to form the \textit{incident context}, which is used for the following profiling, including incident title, symptoms, temporary root causes, and mitigation actions. 

In most cases, OCEs do not follow specific formats to fill in the tickets. For example, the symptoms of tickets could be in various forms, such as textual descriptions, images, tables, runtime logs, and shell scripts. This is because the incidents are very different from each other, and the utmost priority of the OCEs is to mitigate the incident as soon as possible rather than carefully document the tickets. However, these multimodal contents may not be recognized by language models and can add additional noise to the vocabulary. Therefore, we conduct a series of \textit{data cleaning} before feeding the tickets into language models. To deal with that, we first remove the multimodal information, including images and tables from the symptoms. Then, we conduct text preprocessing by removing URLs, HTML tags, and codes using regular expressions and parsers. In this process, we also clean up the text by removing extra spaces, new line marks, and extra braces. 
Finally, we concatenate the selected textual information by adding a brief description for each field to construct the incident context to the language model, which is shown below. The designed format can make the ticket contents more fluent and interpretable to language models, which is beneficial to improve the accuracy~\cite{liu2023prompt_survey}. 
\begin{tcolorbox}[boxsep=1pt,left=2pt,right=2pt,top=3pt,bottom=2pt,width=\linewidth,colback=white!95!black,boxrule=1pt, colbacktitle=white!30!black,toptitle=2pt,bottomtitle=1pt,opacitybacktitle=0.4]
\textbf{Incident context:}
Incident ticket title: [Title]. Symptoms of incidents: [Symptoms]. Identified root cause: [Temporary Root Cause]. Mitigation actions: [Mitigation Actions]
\end{tcolorbox}

\vspace{-2em}
\subsection{Incident Encoder}\label{sec:method-text-encoder}
The incident encoder aims to transform the raw text of incidents into representation vectors, which serve as features for classification. Our approach requires a strong encoder to represent the incident context due to the diverse contents in incident tickets. Therefore, we adopt pretrained language models (PLMs)~\cite{devlin2019bert} to encode incident context, which have demonstrated remarkable ability to understand the semantic meaning of incidents and have proven effective in recent incident understanding tasks~\cite{ahmed2023recommendingrootcause, jin2023oasis}. In this work, we leverage MacBERT~\cite{cui2020macbert} as our text encoder, which is a Transformer encoder model with the same architecture as BERT~\cite{devlin2019bert}. The reason we choose MacBERT over BERT is that MacBERT is an optimized version of BERT trained on the multi-lingual corpus and is capable of processing both Chinese and English. Formally, an input token sequence of incident context is represented as $x=\{\texttt{[CLS]}, x_1, x_2, \dots, x_{n-2}, \texttt{[SEP]}\}$, where $\texttt{[CLS]}$ and $\texttt{[SEP]}$ are two special tokens to indicate the beginning and the end of the sequence. The input sequence is then mapped into input representations and fed into MacBERT to obtain the hidden representations for each token: $\mathbf{X} = \texttt{MacBERT}(x), \mathbf{X} \in \mathbb{R}^{n \times d_{t}}$, where $n$ is the number of tokens in $x$ and $d_{t}$ is the embedding dimension. We use the hidden representation of the first token ($\texttt{[CLS]}$) to represent the whole sequence $\mathbf{x} = \mathbf{X}_{\texttt{[CLS]}}$. 

\subsection{Hierarchy Encoder for Fault Patterns}\label{sec:method-hierarchy-encoder}
The hierarchy encoder aims to transform the fault pattern hierarchy into a series of feature vectors, where each vector represents a node in the hierarchy. In our work, we formulate the fault pattern hierarchy as a Directed Acyclic Graph (DAG) $\; G=(F, E)$, where each node $f_i$ from the node set $F$ contains a node label and description, and the edge set $E$ represents a set of parent-child relations among pairs of nodes, \ie the overall hierarchy. 
Then we use an optimized Graphormer~\cite{ying2021graphormer} to encode the graph, which is the state-of-the-art graph representation architecture based on Transformer layers~\cite{vaswani2017attention} with spatial encoding and edge encoding. 

Our hierarchy encoder first maps the nodes from the graph into a set of input feature vectors. The default Graphormer uses a randomly initialized label embedding of $f_i$ as the input vector $\mathbf{f}_i$. However, in our task, each node contains a clear natural language description of the fault label, which we believe provides fruitful information and can benefit the node representation. Therefore, we adopt the optimized input feature vector of a node to enrich the semantics, which is computed as the sum of label embedding and its description embedding:
\begin{equation}
    \mathbf{f}_i = LabelEmbedding(f_i) + DescriptionEmbedding(f_i)
\end{equation}
The label embedding is randomly initialized and learnable during training with a dimension of $d_t$. The description embedding is computed by a MacBERT encoder using the average of hidden token representations of the description, which also has a size of $d_t$. Specially, we obtain the textual description depending on the node type. For fault patterns in the leaf nodes, we simply concatenate the fault name, instance-level examples, and fault tolerance measures with a brief description for each field. For the fault node in the first four levels, we concatenate its name and description to form the textual input. Finally, we obtain all the $k$ input node feature vectors, which can be stacked as a matrix $\mathbf{F} \in \mathbb{R}^{k\times d_t}$.

After obtaining input node features, the hierarchy encoder injects the parent-child relations to obtain hierarchy-aware node representations with Graphormer architecture:
\begin{equation}
    \mathbf{H} = \texttt{Graphormer}(\mathbf{F})
\end{equation}
Concretely, Graphormer encodes the structural information by spatial encoding and edge encoding. The edge encoding computes the aggregated weight of edges within the path of two nodes, and the spatial encoding measures the distance between two nodes, where both weight matrices are added into the original Query-Key product matrix in the self-attention layer. Here we omit the Graphormer architecture and inherent computation, and please refer to the original paper for the details.

\subsection{Positive Sample Construction}\label{sec:method-positive-data}
A critical pre-step of contrastive learning is to build challenging positive samples that can be used to contrast~\cite{huang2021cosqa, wu2020clear,alzantot2018generating}. In our task, we aim to construct high-quality positive samples for each label considering the taxonomic hierarchy, which can guide models to acquire hierarchy-aware label  representations~\cite{wang2022hgclr} (Section \ref{sec:experiment-rq3} shows the obtained hierarchy-aware representations). The idea of our positive sample construction approach originates from an observation that when text is classified into a certain category, most words are unimportant~\cite{alzantot2018generating}. For example, when an incident description is classified as ``CPU overload'', some keywords such as ``CPU utilization'' or ``exceed 90\%'' provide strong signals while words like users or clusters have fewer impacts. Therefore, the perturbed text that removes some unimportant tokens while keeping the keywords should maintain the original label and can be regarded as a positive sample. Based on this observation, we construct positive samples and utilize fault pattern hierarchy to guide the keyword selection.

We locate the important keywords under a given label by computing the attention weights to the hierarchy-aware label representation and then gather the tokens with larger weights to build the positive samples. Concretely, given the hidden token representations $\mathbf{X} \in \mathbb{R}^{n, d_t}$ of an incident ticket, we first compute the scale-dot attention~\cite{vaswani2017attention} to the node representation $\mathbf{H} \in \mathbb{R}^{k, d_t}$ to obtain the attention weight matrix $A \in \mathbb{R}^{n, k}$, where each element $A_{ij}$ determines the importance of the $i$-th token on a node label $f_j$. After that, we sample important tokens from the importance matrix for a given label $f_j$ to form a positive sample $\hat{x}$. To make the sampling differentiable, we replace the softmax function with gumbel-softmax~\cite{jang2016gumbel} to calculate the probability that $x_i$ is the keyword of class $f_j$ by:
\begin{equation}
    P_{ij} = {gumbel\_softmax}(A_{i1}, A_{i2}, \dots, A_{ik})_j,
\end{equation}
which satisfies $\sum_{j}P_{ij}=1$. Since each incident can have multiple fault pattern labels, we obtain the final importance score of each token $x_i$ by simply adding up the probability of the token regarding its ground-truth label set $f$ as $P_i=\sum_{j\in f}p_{ij}$. The importance scores are used for the final positive sample construction. As one label can have multiple important tokens, we do not transform the probability to one-hot vectors after softmax for discretization. Instead, we keep the tokens for positive samples if their probabilities of being sampled exceed a threshold~$\lambda$. Therefore, the positive sample $\hat{x}$ can be constructed as:
\begin{equation}
    \hat{x} = \{ x_i \; \texttt{if} \; P_i > \lambda \}
\end{equation}
The positive sample $\hat{x}$ is encoded by the same text encoder to obtain the hidden representations: $\hat{\mathbf{X}}=\texttt{MacBERT}(\hat{x})$. The hidden representation of first token ($\textsc{[CLS]}$) is used for sequence representation: $\hat{\mathbf{x}} = \hat{\mathbf{X}}_{\texttt{[CLS]}}$. 

\subsection{Contrastive Learning}\label{sec:method-clr}
After obtaining positive samples, we adopt contrastive learning~\cite{chen2020simclr} to train hierarchy-aware incident and fault pattern representations for better fault pattern profiling. Contrastive learning aims to learn representations by enforcing positive samples to be closer while keeping negative samples further apart. This is achieved by leveraging a contrastive loss function to maximize the similarities of positive samples within the batch and has been proven effective in learning strong representations in many classification tasks~\cite{wei2022clear, qi2023logencoder, shahariar2023contrastiveAPI}. Specially, for each incident sample, we have one positive sample constructed by the method in Section~\ref{sec:method-positive-data}, and $2(N-1)$ negative samples which are all the remaining samples except for $x$ and $\hat{x}$ in the training batch with a batch size of $N$. Finally, we compute the NT-Xent~\cite{chen2020simclr} contrastive loss function of all examples in the batch: 
\begin{equation}
    Loss^{contra} = - \sum_{i=1}^{2N}{\log{\frac{e^{\texttt{cosine}(\mathbf{x}^{(i)},\hat{\mathbf{x}}^{(i)})/\tau}}{\sum_{j=1,j\neq i}^{2N}{e^{\texttt{cosine}(\mathbf{x}^{(i)},\mathbf{x}^{(j)})/\tau}}}}},
\end{equation}
where $\texttt{cosine}(\cdot,\cdot)$ is the cosine similarity between two vectors and $\tau$ is a temperature hyperparameter. 

\vspace{-0.5em}
\subsection{Classification and Training Objective}\label{sec:method-loss}
\paragraph{Classification} The classification module aims to map each incident feature vector into a set of labels. Following previous works~\cite{zhou2020hiagm, wang2022hgclr}, we feed the incident representation $\mathbf{x}$ into a linear classifier with a sigmoid activation function for multi-label classification. The probability on fault labels $f_j$ is computed as: 
\begin{equation}
    p_j = \texttt{sigmoid}(W_c\cdot \mathbf{x} + b_c)_j,\label{eq:cls}
\end{equation}
where $W_c \in \mathbb{R}^{k\times d_h}$ and $b_c \in \mathbb{R}^{d_h}$ are the weights and the bias term. The labels with the probabilities exceeding a certain threshold will be collected as the prediction, which is set to $0.5$ in our work. Notice that we train the incident encoder with hierarchy-guided contrastive learning, which is supposed to be injected with the knowledge of fault pattern hierarchy. Thus, we do not need to incorporate the node representation encoded by hierarchy fault pattern encoder during classification, which is more computationally efficient and effective. 

\paragraph{Training Objective}
During training, we jointly optimize all parameters of our model, including the incident encoder, hierarchy encoder, positive sample construction module, and the classification module. These components work together to categorize fault patterns and learn representations in a contrastive manner. For fault pattern profiling, we employ \textit{weighted binary cross-entropy loss}, a commonly used loss function for multi-label classification. In this context, a weight parameter $\gamma$ is introduced: 
\begin{equation}
\small
    Loss^{cls} = - \sum_{i=1}^{N}\sum_{j=1}^{k}\gamma\dot f_{j}^{(i)}\log{(p_{j}^{(i)})} + (1-f_{j}^{(i)}) \log{(1-p_{j}^{(i)})}\label{eq:cls-loss}.
\end{equation}
The role of $\gamma$ is to mitigate class imbalance and enhance model's sensitivity towards infrequent classes. The objective is to minimize this loss, \ie, to make the predicted probability as close as possible to the true label. 

In Section~\ref{sec:method-positive-data}, we construct positive samples by keeping a few important tokens, which are supposed to maintain the original labels. To train a more robust classifier, we involve the constructed positive samples and compute the classification loss  $\hat{Loss}^{cls}$. Similar to Equation~\ref{eq:cls-loss}, we adopt binary classification loss by substituting $p_j^{(i)}$ to $\hat{p}_j^{(i)}$, where the probability $\hat{p}_j^{(i)}$ can be obtained with the same classification module. 

The final loss function is the combination of classification loss of original samples, classification loss of constructed positive samples, and contrastive learning loss function:
\begin{equation}
    Loss = Loss^{cls} + \hat{Loss}^{cls} + \alpha Loss^{contra},
\end{equation}
where $\alpha$ is a hyperparameter to control the weight of $Loss^{contra}$.

\section{Evaluation}\label{sec:evaluation}
We evaluate our method by answering the following research questions (RQs):
\begin{itemize}[leftmargin=*, topsep=0pt]
    \item RQ1: How effective is \nm in fault pattern profiling?
    \item RQ2: How does hierarchy-guided contrastive learning affect \nm? 
    \item RQ3: How does \nm perform on diverse types of incidents? 
\end{itemize}

\subsection{Experiment Designs}\label{sec:experiment-design}
\paragraph{Dataset}
We collect incident tickets from the incident management system at \cloud, which are created from January 1, 2017, to December 31, 2022. The system is used by a range of service teams such as computing, networking, and storage. As \nm is proposed to assist in postmortem analysis, we conduct filtering to obtain incidents that are ready for postmortem. In particular, we only include tickets from the ``Mitigated'' ones and remove those with all empty contents in the fields of symptom, temporary root cause, and mitigation actions. Finally, we collect 22,560 incidents and 1,463 with annotated fault pattern labels. The number of annotated samples is low since the profiling was done in the postmortem analysis stage, and the postmortem is not done for every incident. 

We use the 1,463 incident tickets that are labeled with fault patterns to form our dataset for evaluation. The dataset is randomly split into training, valuation and test with a portion of 80\%:10\%:10\%. We tune our model on the validation set to search for the best hyperparameters and report the results on the test set.

\paragraph{Implementation Details}
We conduct our experiments on a Linux GPU server with Intel Xeon 2.3GHz CPU and NVIDIA Tesla T4 16G GPU. We implement \nm with Python 3.7.10, PyTorch 1.10.0~\cite{paszke2019pytorch} and transformers 4.2.1~\cite{wolf2019huggingface}. 
For Graphormer, we set the attention head to 8 and feature size $d_t$ to 768, which is the same dimension as the representations produced by the MacBERT encoder $d_h$. The maximum input token length of MacBERT is 512. 
During training, we use Adam~\cite{adam} optimizer with a learning rate of 1e-5 and linear scheduling with 5\% warm-up. We set the training batch size as 8 and train the model for 100 epochs. 
As for hyperparameters, the contrastive loss weight $\alpha$, binary classification weigh $\gamma$ and threshold $\lambda$ are selected by grid search on development set where $\alpha$ is set to $0.1$, $\gamma$ is set to $5$ and $\lambda$ is set to $0.01$. The temperature $\tau$ of the contrastive module is set to 1. 

\subsection{RQ1: How effective is \nm in fault pattern profiling?}\label{sec:experiment-rq1}
\noindent\textbf{Setup.} In this RQ, we evaluate the effectiveness of \nm on the fault pattern profiling task. We compare the whole \nm model against three baseline methods to classify fault patterns. We use the average precision, average recall and F1 score over all examples as the comparing metrics. Specifically, we compare \nm with the three following baseline methods. 
\begin{itemize}[leftmargin=*, topsep=0pt]
\item \textbf{Dense Passage Retriever}~\cite{karpukhin2020dense} (DPR) is the state-of-the-art text matching model, which relates the incident context to fault descriptions in a joint vector space to find the relevant fault patterns. We simply take the top-5 labels as predictions since taxonomy has five levels. The DPR is unaware of the hierarchy.
\item \textbf{MacBERT}~\cite{cui2020macbert} is a multi-label text classifier that feeds the concatenation of incident context and fault descriptions into a MacBERT classifier to obtain relevance scores. It treats taxonomy as flattened labels without considering the structure.
\item \textbf{ChatGLM}~\cite{du2022chatglm} is a bilingual (English and Chinese) large language model with 13B parameters. We tune ChatGLM to classify fault patterns among flatted labels without structure with a parameter-efficient tuning method based on P-Tuning~\cite{liu2022ptuning} at INT4 quantization level. 
\item \textbf{HiAGM}~\cite{zhou2020hiagm} is a state-of-the-art hierarchical text classification model which matches incident context embeddings to fault pattern embeddings encoded by Graph Convolution Networks. HiAGM has leveraged the hierarchical relationships.
\end{itemize}

\begin{table}[t]
  \centering
    \caption{Experiment results of different models.}
\vspace{-1mm}
    \resizebox{0.36\textwidth}{!}{
    \begin{tabular}{lccc}
    \toprule
      \multicolumn{1}{c}{\textbf{Method}} & \multicolumn{1}{l}{\textbf{Precision}} & \multicolumn{1}{l}{\textbf{Recall}} & \multicolumn{1}{l}{\textbf{F1-score}} \\
    \midrule
    Dense Retriever & 48.5 & 61.1 & 54.1 \\
    MacBERT & 58.5 & 61.9 & 60.1 \\
    ChatGLM & 60.0 & 65.2 & 62.5 \\
    HiAGM & 72.1 & 78.2 & 75.1 \\
    \midrule
    \textbf{\nm} & \textbf{76.6} & \textbf{80.1} & \textbf{78.3} \\
    
    \bottomrule
    \end{tabular}%

}
\vspace{-1mm}
  \label{tab:result-overall}%
\end{table}

\noindent\textbf{Results.}
Table \ref{tab:result-overall} shows the performance of different models on fault pattern profiling. We observe that \nm performs substantially better than the other three baseline methods among all metrics, indicating the superiority of \nm in identifying correct fault patterns for incident tickets. In addition, models leveraging the taxonomic structures (\ie \nm and HiAGM) outperform models without considering the structures (\ie Dense Retriever, MacBERT, and ChatGLM) by a large margin. This result demonstrates the importance of injecting hierarchical information to guide models to capture label relationships for better profiling.

\subsection{RQ2: How does hierarchy-guided contrastive learning affect \nm?}\label{sec:experiment-rq2}
\noindent\textbf{Setup.} \nm leverages contrastive learning guided by taxonomic hierarchy to train hierarchical aware incident representation. In this RQ, we examine the effectiveness of hierarchy-guided contrastive learning on the fault pattern profiling task. We first compare the performance of \nm against its variants by removing or replacing different components in \nm. Then, we analyze the label representations of fault patterns to reveal the learnt hierarchy. Specially, we consider the following variants and report the precision, recall, and F1 score on level~\Rmnum{4} for illustration. 
\begin{itemize}[leftmargin=*, topsep=0pt]
\item \textit{r.p.} GCN: We replace Graphormer with Graph Convolutional Network~\cite{kipf2016gcn} (GCN) as the backbone of hierarchy encoder. 
\item \textit{r.p.} GAT: We replace Graphormer with Graph Attention Network~\cite{velivckovic2018gat} (GAT) to encode hierarchy. 
\item \textit{w.o.} Graphormer: We remove then Graphormer encoder and directly use the node vectors to guide positive sample construction.
\item \textit{w.o.} description embedding: We remove description embedded in the hierarchy encoder and only use the label embedding.
\item \textit{w.o.} contrastive loss: We remove contrastive loss function.
\item \textit{w.o.} augmented samples loss: We remove the classification loss function for augmented positive samples. 
\item \textit{w.o.} whole contrastive module: We remove the entire hierarchy-guided contrastive learning module and only leverage the incident encoder and classification module for profiling.
\end{itemize}

\begin{table}[t]
  \centering
    \caption{Experiment results of different models.}
    \resizebox{0.42\textwidth}{!}{
    \begin{tabular}{lcccc}
    \toprule
      \multicolumn{1}{l}{\textbf{Method}} & \multicolumn{1}{l}{\textbf{Precision}} & \multicolumn{1}{l}{\textbf{Recall}} & \multicolumn{1}{l}{\textbf{F1-score}} \\
    \midrule
    \textbf{\nm} & \textbf{76.6} & \textbf{80.1} & \textbf{78.3} \\
    \textit{-r.p.} GCN & 71.4 & 74.2 & 72.8 \\
    \textit{-r.p.} GAT & 71.9 & 74.8 & 73.3 \\
    \textit{-w.o.} description embedding & 72.8 & 75.1 & 74.0 \\
    \textit{-w.o.} Graphormer & 66.2 & 71.8 & 68.9 \\
    \textit{-w.o.} contrastive loss  & 67.2 & 75.5 & 71.3 \\
    \textit{-w.o.} augmented samples loss  & 53.4 & 64.4 & 58.4 \\
    \textit{-w.o.} whole contrastive module  & 50.6 & 59.5 & 54.7 \\
    \bottomrule
    \end{tabular}%

}
  \label{tab:result-ablation-clr}%
\end{table}

\noindent\textbf{Results.} The results in Table \ref{tab:result-ablation-clr} shows that: 
(1) Graphormer exhibits superior performance compared to both GCN and GAT hierarchy encoders in this task. This can be attributed to the fact that GCN and GAT encode local structures, which only perform convolutions or attentions on neighboring nodes, whereas Graphormer employs global attention, allowing each node to attend to all others in the graph. This global attention mechanism is more effective in encoding hierarchy.
(2) Removing the description embedding from the node embedding leads to a decrease in performance for \nm. This result highlights the significance of fault descriptions, as they provide additional semantic information that helps in learning more effective representations.
(3) When the Graphormer encoder is entirely discarded, there is a significant drop in performance, indicating the usefulness of hierarchy in providing organized label relations that guide fault pattern profiling.
(4) Without the training objectives of contrastive loss or classification loss of positive samples, \nm performs poorly. This indicates that both the positive sample construction and the contrastive learning framework contribute to \nm. The positive samples are useful even without contrastive learning, and contrastive learning can further enhance the model by constraining the vector space. 
(5) Removing the entire hierarchy-guided contrastive learning module results in a substantial accuracy decline for \nm. This finding again confirms the effectiveness of our method in hierarchical fault pattern profiling. 

\begin{figure}[t]
    \centering
    \includegraphics[width=0.92\columnwidth]{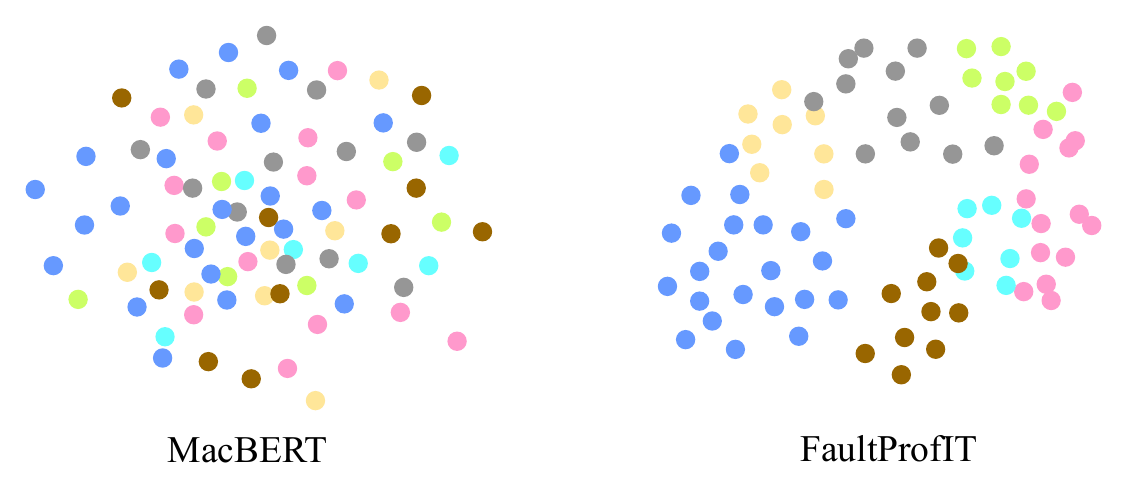}
    \vspace{-1em}
    \caption{Visualization of the fault pattern representations.}
    \label{fig:tsne-fp-distribution}
    \vspace{-1em}
\end{figure}

\noindent\textbf{Visualization.} To investigate the encoding of the taxonomic hierarchy, we visually analyze the distributions of fault pattern embeddings. Specifically, we consider the weight matrix~$W_c \in \mathbb{R}^{k \times d_h}$ in Equation~\ref{eq:cls} as fault pattern representations. where each row represents a node in the taxonomy. We employ the T-SNE algorithm~\cite{van2008tsne} with default parameters to project the high-dimensional vectors into a two-dimensional space. The resulting points are plotted in Figure~\ref{fig:tsne-fp-distribution}, where points with the same color correspond to fault patterns from the same parent node. For comparison, we also visualize the embeddings produced by MacBERT. Our observations reveal that the fault pattern embeddings of MacBERT are scattered, while the embeddings of \nm exhibit clustering based on the parent nodes. This behavior arises from the fact that the representation of a label and its parent is trained to be similar, as they are classified simultaneously. Consequently, if the hierarchy is incorporated into the text representation, labels sharing the same parent should possess more similar representations compared to those with different parents.

\subsection{RQ3: How does \nm perform on diverse types of incidents?}\label{sec:experiment-rq3}
\noindent\textbf{Setup.} Developing a strong fault pattern profiler necessitates a substantial volume of training data. To accommodate this requirement, we utilize a number of historical incident tickets collected from a variety of services and encompassing a range of severity levels. In this RQ, we explore the performance of \nm across diverse incident types. To this end, we partition the testset into several subsets based on their severity and the services they impact. Given the existence of over 30 distinct services, we categorize them into five primary groups: \textit{Infrastructure}, \textit{Computing}, \textit{Networking}, \textit{Storage}, and \textit{Others}. 

\begin{figure}[t]
    \centering
    \includegraphics[width=0.99\columnwidth]{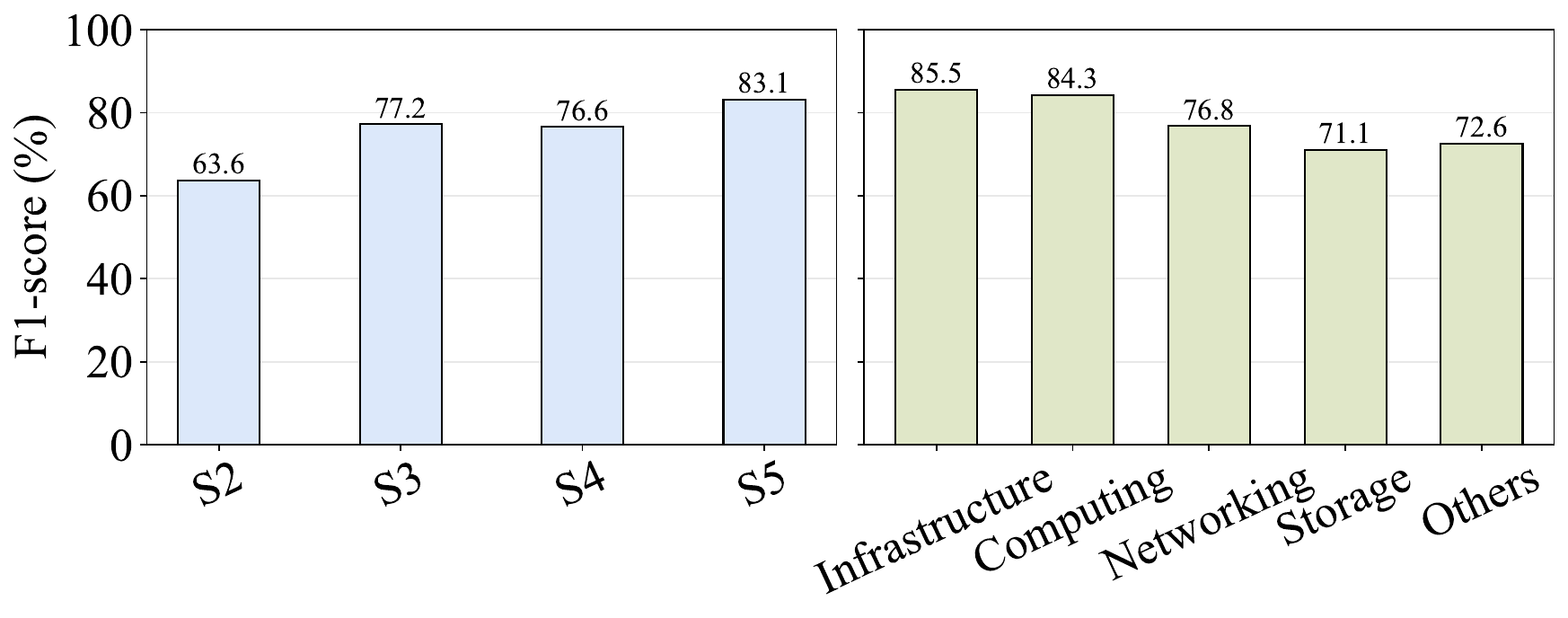}
    \vspace{-1em}
    \caption{Results of \nm for incidents of different severities and services.}
    \label{fig:RQ3}
    \vspace{-1em}
\end{figure}

\noindent\textbf{Results.} Figure~\ref{fig:RQ3} shows the F1-score of \nm across various incident types. From the left segment of Figure~\ref{fig:RQ3}, we can find that \nm exhibits superior performance on incidents with less severity, especially those of S5 level, compared to more severe ones. This could be attributed to the increased complexity in diagnosing severe incidents, which often involve extended incident contexts and a greater number of affected components, thereby posing a greater challenge for \nm to identify. These findings further confirm the viability of automated fault pattern profiling within cloud systems, given the higher frequency of less severe incident tickets and fewer human resources allocated for analysing these tickets, compared to those of higher severity.

The right segment of Figure~\ref{fig:RQ3} indicates that the accuracy varies across services. Incidents impacting services within the \textit{Infrastructure} and \textit{Computing} categories yield a relatively high F1-score. Conversely, incidents affecting services within \textit{Storage} or \textit{Others} categories demonstrate lower accuracy. Different from other services, the \textit{Infrastructure} and \textit{Computing} services mainly experience faults within clusters and hosts, which comprise servers and hardware. These incidents often exhibit explicit descriptive signals such as ``server'' or ``datacenter'', thereby facilitating easier classification. The lower F1-score associated with \textit{Storage} services can be attributed to the smaller number of incidents within this category present in our dataset, as the service is more robust and produces fewer failures. The low accuracy of the \textit{Others} category is because it consists of multiple services, each with diverse phenomenons and mitigation methods. As a result, \nm encounters difficulties in discerning common semantic patterns within this category, leading to a higher rate of erroneous predictions.

\section{Deployment Experience}
In this section, we share our experience of deploying \nm in product X, a cloud reliability analysis system at \cloud. This system provides an extensive array of centralized analytic functionalities for incident management, including tracing, retrieval, analysis, and modeling.
These functionalities are tailored to support a variety of service teams and engineers in managing their production incidents and identifying reliability issues. 
At \cloud, the reliability team conducts incident postmortem analysis with product X. One of the major outcomes is the profiling of fault patterns, which are subsequently utilized for trend analysis and vulnerability identification. In most cases, the reliability team does not conduct postmortem analysis once the incident is mitigated. Instead, they perform analysis periodically, such as weekly or monthly, selecting a set of severe incidents to investigate and profile fault patterns.

Traditionally, the process of fault pattern profiling relied on manual labeling. However, with the introduction of new service products and the increasing number of customers, engineers found it increasingly challenging to analyze emerging incidents. To reduce efforts, reliability engineers prioritized severe incidents (\ie S1, S2, and S3) for profiling and proposed product improvement suggestions based on the profiling results.
However, such practices neglect the incidents with less severity. Even though system updates were frequently released to improve reliability, the number of minor incidents continued to increase. 
Although such incidents did not cause severe impacts thanks to the fault tolerance measures, specific customers suffered from occasional performance degradation or network interruption, affecting customer experience and causing complaints.
Therefore, the integration of automated tools for fault pattern profiling is essential to improve both the efficiency and comprehensiveness of the analysis process.

To achieve this goal, we have integrated \nm into product X. Currently, \emph{10,000+} of incidents from \textit{30+} cloud services (including historical incidents) have been analyzed by \nm for fault pattern profiling over a \textit{six-month} period. Concretely, we provide an API interface in Product X. Engineers can invoke the API to call \nm, which automatically analyzes incident tickets to profile fault patterns. When the API is invoked, the unanalyzed tickets are sent to the server. Subsequently, \nm conducts data preprocessing and predicts the labels in a batch manner. Once the prediction is completed, the profiled fault patterns are returned and visualized on the frontend of product X.

\begin{figure}[t]
    \centering
    \includegraphics[width=0.92\columnwidth]{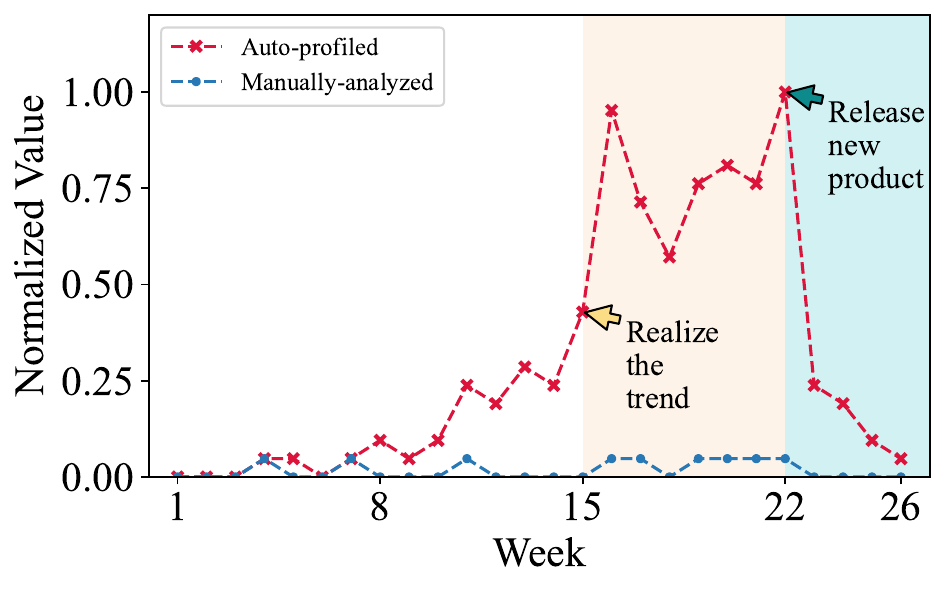}
    \vspace{-1em}
    \caption{An example of trends analysis for the \textit{memory overload} fault pattern with \nm.}
    \label{fig:industry}
    \vspace{-1em}
\end{figure}

To show how the predicted fault patterns are utilized in \cloud, we collected incidents created over the most recent 26 weeks after the deployment. Their fault patterns are automatically profiled by \nm on a weekly basis. We present an example trend of the fault pattern named \textit{memory overload} from a computing service, which indicates that the system's memory usage had exceeded its capacity.
Figure~\ref{fig:industry} shows the trends of incidents with memory overload from two resources\footnote{Due to confidential reasons, we present a normalized occurrence number to reflect the trends.}. The blue line represents manually analyzed postmortem reports, which are conducted only on severe incidents, and the red line represents automatically profiled incidents. In the first 10 weeks, the occurrence of incidents of the fault pattern remained at a low level. Then from week 10 onwards, the number of auto-profiled incidents began to rise, reaching a peak at week 16. Without the integration of \nm, engineers would be unaware of the faults their service was undergoing, as the memory overload is not a severe fault to cause large impacts and thus they are not analyzed in the postmortem. However, with \nm, they were alerted to an increasing number of memory overload incidents at week 15. Therefore, the team initiated a set of actions to investigate the overload issues within the service. During the two-month investigation, engineers conducted numerous experiments to test the system, identify weaknesses, and fix the defects. Finally, they released a new version of the service at week 22, and this fault pattern began to decrease and eventually returned to a low level by week 26.

\section{Related Works}
With years of efforts, researchers have conducted numerous studies~\cite{huang2017gray,chen2020towards,gunawi2016does,liu2019bugs} and proposed many automatic approaches~\cite{lee2023maat,li2021fogofwar,gu2020efficient,gu2020efficient2,li2022intelligent,zhang2021onion,lee2023maat,chen2021graph} on cloud incidents management. Among these works, Gunawi et al.~\cite{gunawi2016does} discussed why incidents still take place in cloud systems by analyzing public incident reports of popular cloud services. Chen et al.~\cite{chen2020towards} presented a comprehensive study on how incidents are managed in Microsoft Azure. 

Timely and accurate incident detection can facilitate the quick response of engineers, accelerating the procedures involved in incident management. By analyzing cloud system service behaviors, Warden~\cite{li2021fogofwar} was proposed to analyze system-wide alerting signals from a global view for proactive incident detection. To avoid the flooding issue reports, MID~\cite{gu2020efficient} was proposed to identify incidents from large-amount, multi-dimensional issue reports.

Once the incidents are detected, diagnosis and Root Cause Analysis (RCA) are conducted to obtain comprehensive information that aids in follow-up triage and choosing effective mitigation strategies. Onion~\cite{zhang2021onion} localizes the incident-indicating logs from the incident context, where a contrast analysis is then performed to accurately find out a few lines of root cause related log. ESRO~\cite{chakraborty2023esro} constructs a unified graph of alerts and incident reports to recommend root causes and remediation steps. iPACK~\cite{liu2023incident} and LinkCM ~\cite{gu2020efficient2} were proposed to link and aggregate duplicate incidents by fusing the failure information between the customer-reported tickets and the machine-generated incidents. To avoid excessive aggregation of incidents, HALO~\cite{zhang2021halo} was proposed to localize the fault to a proper granularity, which usually suffers from improper aggregation level of incidents for further diagnosis and triage. Chen et al.~\cite{chen2019triage} conducted an empirical study about incident triage in Microsoft Azure and further proposed DeepCT~\cite{chen2019continuous} to automate continuous incident triage for further incident mitigation.

After the mitigation of incidents, postmortem analyses are conducted to provide fruitful insights and experiences. This knowledge can assist in profiling the fault, contributing significantly to the enhancement of the system's stability and response speed. Shetty et al.~\cite{shetty2022softner} conducted an empirical study on hundreds of high severity incidents postmortems in a large-scale cloud service and provided guidance on how to tackle future incidents. AutoARTS~\cite{dogga2023autoarts} was proposed to label incident root causes by analyzing potential contributing factors with knowledge gained from incident postmortems.

The rise of Large Language Model (LLM) has brought new opportunities to the field of intelligent incident management. With intrinsic domain knowledge, LLM can diagnose and interpret incidents like on-call engineers (OCEs). Ahmed et al.~\cite{ahmed2023recommendingrootcause} effectively fine-tuned LLMs to suggest the root cause and mitigation strategies for cloud incidents, combining both external domain expertise and internal pre-trained model knowledge. Moreover, RCACopilot~\cite{chen2023empowering} was proposed to summarize the incidents and predict the incident’s root cause with generated explanations by employing LLMs.

Our work focuses on the postmortem analysis phase of incident management. We distinguish from existing works by developing an automated approach to profile fault patterns based on incident tickets, which can handle emerging and less severe incidents. Our work can provide a comprehensive view of a range of incidents and improve the efficiency of reliability engineers.
\section{Conclusion}
Fault pattern profiling is an important task of incident postmortem analysis in large-scale cloud systems. To support consistent and large-scale fault pattern profiling, we introduce \nm, an automated approach that leverages hierarchical text classification. \nm takes the textual incident context as input and applies language models to predict fault pattern labels. To inject hierarchy information into the taxonomy and mitigate the data insufficiency problem, we employ hierarchy-guided contrastive learning to enhance the incident representations. We evaluate our approach to the production incidents of \cloud, a top-tier global cloud provider. Experimental results demonstrate the high F1-score achieved by \nm and the effectiveness of hierarchy-guided contrastive learning. Furthermore, we have deployed \nm at the reliability analysis platform of \cloud for a duration of six months, gaining valuable insights and experience from the deployment. 

\section*{Acknowledgement}
We sincerely thank the anonymous reviewers for their constructive comments and suggestions. The work described in this paper was supported by the Research Grants Council of the Hong Kong Special Administrative Region, China (No. CUHK 14206921 of the General Research Fund).


\balance
\bibliographystyle{ACM-Reference-Format}
\bibliography{icse-seip24}

\end{document}